\documentclass{aa}

\usepackage{graphicx}
\usepackage{txfonts}

\usepackage{float}

\usepackage{natbib}                                                     
\begin{document} 
\titlerunning{The role of bars in triggering AGN galaxies}
 \authorrunning{Marels et al.}

   \title{The role of bars in triggering active galactic nuclei galaxies}

   \author{V. Marels\inst{1},  V. Mesa\inst{1,2,3}, M. Jaque Arancibia\inst{1}, S. Alonso\inst{4}, G. Coldwell\inst{4}, G. Damke\inst{5}, V. Contreras Rojas\inst{1} }

   \institute{
Departamento de Astronomía, Universidad de La Serena. Av. Raul Bitrán, La Serena, Chile.\\
\email{valeria.alvarezv@userena.cl}
\and
Association of Universities for Research in Astronomy (AURA)
\and
Grupo de Astrofísica Extragaláctica-IANIGLA, CONICET, Universidad Nacional de Cuyo (UNCuyo), Gobierno de Mendoza
\and
Departamento de Geofísica y Astronomía, CONICET, Facultad de Ciencias Exactas, Físicas y Naturales, Universidad Nacional de
San Juan, Av. Ignacio de la Roza 590 (O), J5402DCS, Rivadavia, San Juan, Argentina
\and
Cerro Tololo Interamerican Observatory, NSF’s NOIRLab, Av. J. Cisternas 1500 N, La Serena, Chile
}

   \date{Received 1 April 2025; accepted 19 May 2025}

  \abstract

   {Bars are considered an efficient mechanism for transporting gas toward the central regions of galaxies, potentially enhancing nuclear activity. However, the extent to which bars influence active galactic nuclei (AGNs) and whether their efficiency varies with environment remain open questions.}
   {In this study, we aim to quantify the role of bars in triggering AGNs by comparing the AGN fraction in barred and non-barred galaxies across different environments.}
   {We constructed a sample of barred and non-barred galaxies from the Galaxy Zoo DECaLS catalog, ensuring a control selection where both samples share similar distributions in stellar mass, redshift, magnitude, concentration index and the local density parameter. AGNs were identified using spectroscopic data from Sloan Digital Sky Survey in order to obtain a final sample of barred AGNs galaxies (1330) and a control sample of unbarred AGNs (1651). We employ the [OIII]$\lambda$5007 luminosity ($Lum[OIII]$) and the accretion rate parameter $\cal{R}$ parameter as indicators of nuclear activity, based on these parameters, we applied specific criteria to distinguish between powerful and weak AGNs, allowing for a more precise assessment of the potential impact of bars on the supermassive black hole.}
   {Our analysis reveals that barred galaxies tend to host a higher fraction of powerful AGNs compared to unbarred galaxies. From the analysis of Lum[OIII] we find that galaxies with higher nuclear activity tend to be massive, blue and with young stellar populations. In addition, we observed a slight tendency for barred galaxies to host less massive black holes, which were found to accrete matter more efficiently.
   The classification analysis of strong and weak bars indicates that galaxies with a more prominent bars exhibit higher nuclear activity. Furthermore, we study the environmental dependence of this trend. Although no significant differences between strong and weak bars were found in intermediate density environments, we observe a distinction in both low and high density environments, where galaxies with strong bars exhibit higher AGN activity.}

   {}


   \maketitle

\section{Introduction}

The main fueling mechanism of the supermassive black holes (SBMHs) are related to dynamical perturbations within the host galaxy, which drive significant amounts of gas toward the central regions of the galaxy, causing intense activity around the nuclei \citep{LyndenBell1969}. This is the most widely accepted hypothesis about Active Galactic Nuclei \citep[AGN;][]{Richstone1998}. The question of what mechanism is responsible for triggering this activity around a SMBH still has no clear answer, which is fundamental to understanding the co-evolution between AGN and its host galaxy. 

In this context, the morphology play an important role due to the structure will define how the galaxy evolves and how internal mechanisms facilitate the inflow of gas towards the nucleus.
Taking this into account, several authors agree that it is necessary to consider galactic bars as a possible processes for torquing material of galactic centers \citep{Mihos1996,Athanassoula2003,Alonso2013,Kim2020,Garland2024}. The surrounding gas and dust under these conditions can precipitate towards the nucleus activating the black hole activity or increasing it.

Bars are very common structures in some disk-type galaxies composed by stars formed due to gravitational instabilities in the galactic disk \citep{Sellwood2014}. The fraction of bars in disc galaxies in the local Universe is as high as 70\%-80\% in infrared wavelengths \citep{Eskridge2000, Menendez2007}. However, in optical wavelengths, the observed bar fraction is typically lower, around 30\% \citep{Masters2011}. They are expected to exert an essential influence in the dynamical evolution of their hosts and even on the episodes of star formation \citep{Heckman1980,Ellison2011a}, in addition to causing modifications in the galactic structure \citep{Buta1996}, in the chemical composition of the gas \citep{Martin1994} and in some cases producing nuclear starburst \citep{Gadotti2001}. The idea that they can drive a flux of matter to the central regions of the galaxy and contribute to its evolution \citep{CombesElmegreen1993}, comes about due the gas clouds within galaxies undergo collisions with the edges of the bar \citep{Shlosman1990} leading to a redistribution of gas and stellar component increasing the probability that the galaxy hosts an AGN. It should also be noted that it is possible that the bars explain the formation of the bulges \citep{Kormendy2004}. Considering this, \cite{Coelho2011} found that for low-mass bulges, the occurrence of AGN increases in barred galaxies compared to non-barred galaxies.

There is a large amount of observational evidence to support the fact that bars can enhance nuclear central activity, however, there is still no complete agreement on the relationship between these phenomena. Firstly, \cite{Knapen2000} and \cite{Laine2002} exhibit increased bar fractions for active nuclei host galaxies compared to non-AGNs. More recent studies such as \cite{Oh2012} show that late-type barred AGNs have higher values of nuclear activity with less-massive black holes and bluer colors. \cite{Alonso2013} found a higher fraction of powerful nuclear activity in isolated barred AGN compared with non-barred AGNs. Also, they found that these galaxies show a higher fraction of accretion rates in comparison to the control sample, results that are corroborated again in \cite{Alonso2018} and \cite{Alonso2024}, which also includes a sample of pairs of galaxies in order to investigate whether interactions also generate increases in nuclear activity. Furthermore, \cite{Alonso2014} studied barred AGNs in different environments, such as groups and clusters, and found an increase in nuclear activity in barred galaxies that were located in higher-density environments. While \cite{Lokas2016} states that the formation of the bars is strongly influenced by the environment in which it is found, \cite{Aguerri2023} explains that it is actually the size of the galaxy that is affected by the environment and consequently the size of the bar adapts to the galaxy. Therefore, the study of internal phenomena such as AGN activity and the presence of a bar in various environments remain a topic of debate necessary to advance in the understanding of a general evolutionary scenario. 

On the other hand, studies such as \cite{Lee2012} and \cite{Cheung2015} found no clear evidence of a close relationship between the presence of a bar and the occurrence or power of an AGN. This discrepancy of results can have several factors that can range from the volume of data used in either barred or active galaxy samples and the precision of their characteristics to the definition used when classifying AGNs in the Baldwin, Phillips and Terlevich diagram (hereafter BPT) \citep{BPT} or emission-line diagrams \citep{Veilleux1987}, as well as possible differences in parameters such as the mass range used and the redshift. \cite{Galloway2015} used galaxies from SDSS and Galaxy Zoo 2 and found that the bar fraction among AGN host galaxies was higher than that among star-forming galaxies. However, there was no evidence that the accretion rate depended on the presence of a bar. Recently \cite{SilvaLima2022} reported that barred galaxies hosted a higher amount of AGN compared to unbarred galaxies and the accretion rate is higher in barred galaxies, but only when different M- $\sigma_*$ relations are used to estimate the black hole mass in barred and unbarred galaxies. They found no correlation between activity power and bar strength.

Simulations have analyzed in part the correlation between an AGN and the properties of its host. \cite{Piner1995} have shown that the gravitational potential of a large-scale bar induces the entry of gas into the center and can provide fuel to initiate or increase nuclear activity. More recently, \cite{Irodotou2022} used AURIGA simulations which show an anti-correlation between the ejective nature of AGN feedback and bar strength.

In this work, we investigate the impact of bars on AGN activity by comparing the properties of barred and unbarred galaxies to assess potential differences between the two samples. This analysis is based on high-resolution images from the Dark Energy Camera Legacy Survey (hereafter DECaLS, \cite{Dey2019}), which provide exceptional depth and morphological detail, enabling precise identification and classification of bars. The superior quality of DECaLS data in terms of spatial resolution and imaging depth is crucial for distinguishing bar structures, allowing us to categorize them by strength and explore their correlation with AGN activity more robustly. This study will also account for environmental factors, comparing barred and unbarred galaxies across various environments, including high, medium and low density regions, to explore how external influences may modulate the relationship between bars and AGN fueling.\\ 

This paper is structured as follows: We present the most relevant features of our data in Section 2, including the AGN catalog and Galaxy Zoo classifications. Section 3 describes the construction of the barred AGN galaxy catalog and the selection criteria for the control sample. In Section 4, we compare the properties of AGN host galaxies, analyzing the influence of bars — distinguishing between strong and weak bars — and the effect of the environment. Finally, in Section 5, we summarize the key results and main conclusions.

The cosmology adopted is $\Omega = 0.3$, $\Omega_{\lambda} = 0.7$ and $H_0 = 100km s^{-1}Mpc$.

\section{Data}

This work is based on morphological data selected from posteriors of Galaxy Zoo DECaLS\footnote{https://data.galaxyzoo.org} \citep[GZ;][]{Walmsley2022}, a citizen science project where volunteers visually classify galaxies based on their morphology. These classifications are then used to train a Bayesian convolutional neural network, wich automates the classification of the remaining galaxies. DECaLS is designed to be one of the deepest surveys to date (r=23.6), with the objective of detecting fainter objects and providing more detailed images over a total area of 9,000 square degrees of the sky, primarily in equatorial regions. Uses images obtained from Dark Energy Camera (DECam), particularly powerful for wide-field observations with a field of view of three square degrees, mounted on the 4-meter Blanco telescope at Cerro Tololo. GZ DECaLS classifies in detail a total of 314,000 galaxies into various categories, which include the presence of a bar, galaxy orientation, bulge prominence, number of arms, etc.

The level of detail provided by GZ DECaLS, allows us to reveal internal structures of galaxies, such as the bars and possible bridges that indicate galaxy interactions that were not visible before. In comparison with the Sloan Digital Sky Survey \citep[SDSS;][]{York2000}, which reaches a magnitude of r=22.2, DECaLS offers deeper imaging and improved spatial resolution, facilitating more accurate bar identification, particularly through visual inspection. The SDSS is one of the largest galaxy survey at the present, covering approximately one-quarter of the celestial sphere and provides spectra for more than a million objects. SDSS Data Release 7 \citep{Abazajian2009} includes 11,663 $deg^2$ of imaging data and contains five-band photometry (u, g, r, i and z) for 357 million distinct objects.\\ 
    \begin{figure}[h!]
   \centering
   \includegraphics[scale=0.34]{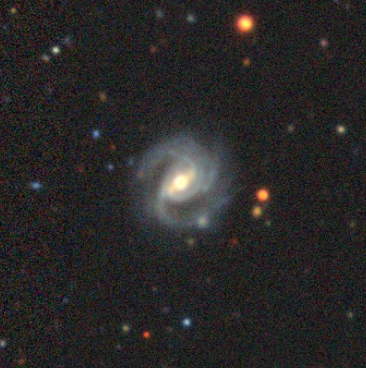}
   \includegraphics[scale=0.325]{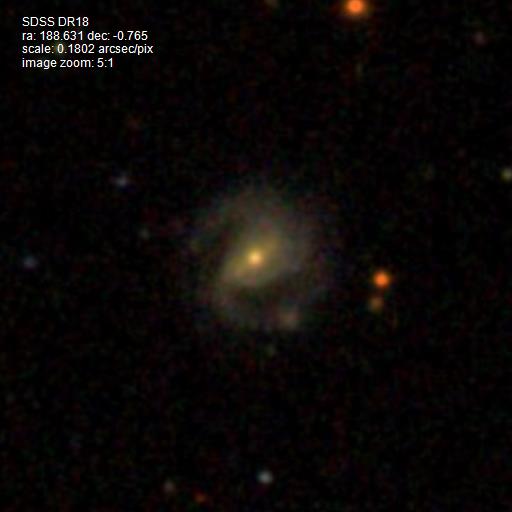}
    \includegraphics[scale=0.34]{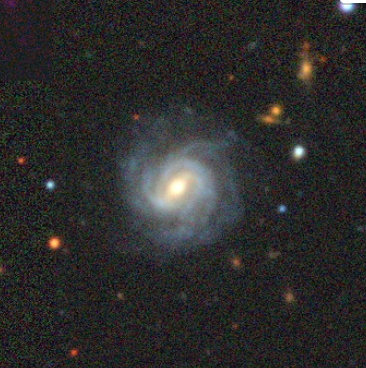}
   \includegraphics[scale=0.325]{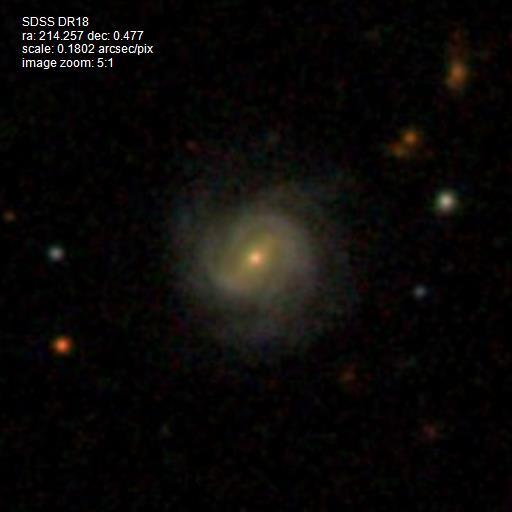}
       \includegraphics[scale=0.34]{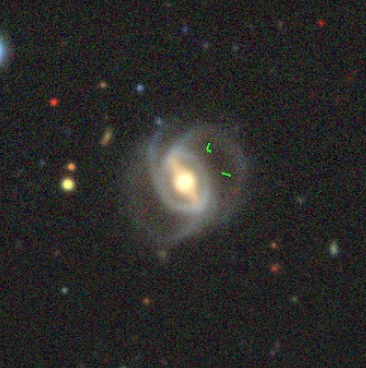}
   \includegraphics[scale=0.325]{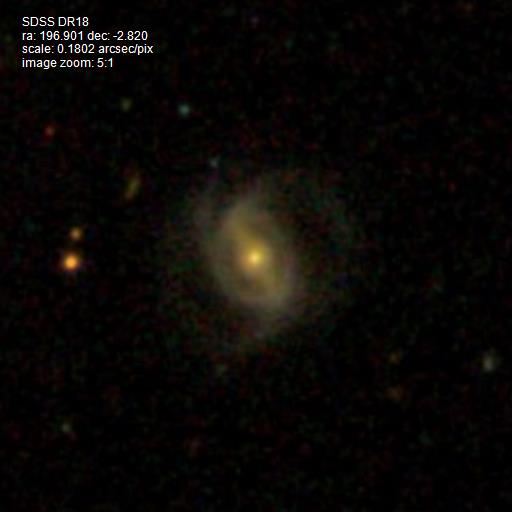}
      \caption{Examples of barred galaxies from DECaLS (left panels) and SDSS (right panels).}
         \label{examples}
   \end{figure}
\\

In this version of Galaxy Zoo, unlike previous iterations based on SDSS data (see Fig. \ref{examples}), the questions were adapted to yield more characteristic classifications of the galaxies. For the first time, automatic classifications were provided, with both the survey depth and the weighting of volunteer votes playing a key role in the model training process.

For AGNs sample, we use the photometric and spectroscopic data from the SDSS Data Release 7 \citep[][SDSS-DR7]{sdss7}. The main galaxy sample consist of approximately 700.000 galaxies with both spectroscopic and photometric measurements. It is a magnitude-limited spectroscopic sample, defined by a Petrosian magnitud limit of $r<17.7$. Most galaxies in this sample lie within the redshift range $0<z<0.25$ \citep{Strauss2002}.

The AGN galaxies were selected using publicly available emission-line flux measurements. The methodology employed to derive these measurements is described in detail by \cite{Tremonti2004}. The emission-line fluxes were corrected for reddening using the Balmer decrement and the attenuation curve from \cite{Calzetti2000}. The signal-to-noise ratio of each line was recalculated using adjusted errors based on the uncertainty estimates provided by the MPA/JHU\footnote{Avaible at http://www.mpa-garching.mpg.de/SDSS/DR7/} catalog. The AGN sample was selected using the BPT diagram. This diagram helps us to distinguish between galaxies whose ionization is dominated by star formation, those dominated by AGN activity. Using relation between specific emissions lines determined by their sensitivity to the conditions of ionized gas and are independent to reddening. We considered the relation between spectral lines [OIII]5007, H$\beta$, [NII]6583 and H$\alpha$. The BPT diagram allows to divide in Seyfert, LINER \citep{Schawinski2007} and composite \citep{Kewley2001}. For this work, we considered all galaxies lying above the \cite{Kauffmann2003} demarcation line, without distinguishing between AGN subtypes. This selection resulted in a sample of approximately 86.000 galaxies classified as AGN.

The procedures to derive the galaxy physical properties are described by \cite{Kauffmann2003,Blanton2005,Salim2007}. These data are provided by MPA/JHU and the NYU\footnote{http://sdss.physics.nyu.edu/vagc/}, all accesible in the SDSS spectroscopic database, that includes emission lines fluxes, stellar masses, star formation rate (SFR) and $D_n4000$ parameter as an indicator of the age of stellar populations. From this catalog we use the star formation rate normalized to the total mass in stars estimated from the SDSS fiber, $log(SFR/M_*)$, taken from \cite{Brinchmann2004}. In addition, we use the total stellar masses $Log(M*/M\odot)$ calculated using the Bayesian methodology, and model grids described in \cite{Kauffmann2003}. We have adopted \cite{Balogh1999} definition of  $D_n(4000)$  as the ratio of the average flux densities in the narrow continuum bands (3850-3950 \AA \  and 4000-4100 \AA). The color used in this work were computed by subtracting the k-corrected \citep{Blanton2007} absolute magnitudes in the r and u bands from the SDSS DR7 catalog, calculated using Petrosian magnitudes converted to the AB system. Finally, to discriminate between bulge and disc-types galaxies, we use the concentration index C\footnote{$C=r90/r50$ is the ratio of Petrosian 90 \%- 50\% r-band light radii} \citep{Abraham1994}, a well tested morphological classification parameter  \citep{Strateva2001}. We obtained  all data  catalogs  through  SQL  queries  in  CasJobs\footnote{ http://skyserver.sdss.org/casjobs/}.

\section{Samples}

\subsection{AGN-barred galaxy catalog}
\label{agn_sample}

For this analysis, we obtained a sample of barred spiral galaxies from GZ DECaLS catalog. We selected galaxies classified as a spiral galaxies with the presence of a bar according to GZ DECaLS, using a probability threshold greater than 0.7 ( $p_{feature-or-disk}>0.7$, $p_{spiral}>0.7$,$p_{strong-bar+weak-bar}>0.7$). This selection ensures a sample with sufficient for statistical analysis while maintaining a high level of refinement, minimizing the inclusion of unbarred or non-spiral galaxies. We also restricted barred spiral galaxies to face-on orientation ($p_{disk-edge-on-no}>0.7$), which favors classification based on visual inspection. In addition, we restricted the sample to galaxies with a redshift lower than 0.1 ($z<0.1$). In order obtain AGNs barred galaxies, we cross-correlated this sample with AGN galaxies from \cite{Coldwell2017} whose catalog contains a total of 86.134 of AGNs sorted by BPT. We found 1330 barred-AGN galaxies with absolute magnitudes in r-band $-23 mag. <M_r< -19 mag.$  and stellar masses $10^{9.5} <M*< 10^{12}$.  Additionally these galaxies were divided in strong and weak bar by the classification provided by GZ DECaLS. A strong bar typically has a much longer and brighter, clearly defined shape whose presence is observed more easily. On the other hand, the weak bars are less “obvious'' \citep{Vaucouleurs1963}, short and not very luminous, which makes them difficult to detect during a visual inspection and can even be said to have a minor impact on the global structure and dynamics of the galaxy. We define a galaxy as strongly barred if $p_{strong-bar}>p_{weak-bar}$ and weakly barred if $p_{strong-bar}<p_{weak-bar}$. More details in Section \ref{Types_bars}.
\\

\subsection{Control sample}
In order to understand the impact of bars on the central nuclear activity, we constructed an appropriate control sample (CS) of unbarred spiral AGNs. We initially selected these galaxies from GZ DECaLS catalog, applying the same criteria as ins section \ref{agn_sample}, requiring a probability ($p_{strong-bar+weak-bar}<0.3$) and a $z<0.1$ and then crossmatched this morphological sample with the AGNs catalog. However, the number of unbarred AGN galaxies obtained from GZ DECaLS was significantly lower (\string~200) than that of barred AGNs, making the control sample insufficient for a robust comparison. To mitigate this issue, we supplemented the control sample with unbarred AGNs from the Galaxy Zoo 2 \citep[GZ2;][]{Willett2013} catalog, ensuring a more balanced dataset. To minimize contamination from potentially barred galaxies misclassified due to SDSS image quality, all GZ2 galaxies included in the control sample were visually inspected using DECaLS imaging to confirm the absence of a bar. While GZ DECaLS was prioritized due to its improved bar identification, the inclusion of GZ2 allowed us to obtain a statistically reliable control sample. 

Using SDSS mock galaxy catalogs based on the Millennium Simulation, \cite{Perez2009} demonstrated that an appropriate control sample for galaxy pairs must be selected to match at least in redshift, stellar mass, and local density environment. These criteria are also applicable when constructing control samples of barred galaxies, aimed at analyzing their properties in comparison to unbarred ones.

Following this approach, galaxies in the control sample were randomly selected to match the barred galaxies in the distribution of three key parameters: redshift ($z$), stellar mass ($M*$), and local density environment ($\Sigma_5$). Additionally, we extended the comparison to include two further parameters, absolute magnitude ($M_r$) and concentration ($C$).

The local density parameter was defined through the project distances $r_p$ to the fifth nearest neighbor, \textbf{$\Sigma_5 = 5/(\pi r_{p}^2)$}, with luminosities brighter than $M_r < -20.5$ \citep{Balogh2004} and radial velocity difference of less than 1000$km/s$. $\Sigma_5$ provides a adequate measure of the local density of the galaxies. 
This selection ensures a comparable sample for analyzing the local density environment of the galaxies and estimating the conditions in which they reside.

       \begin{figure}[h!]
   \centering
   \includegraphics[scale=0.3]{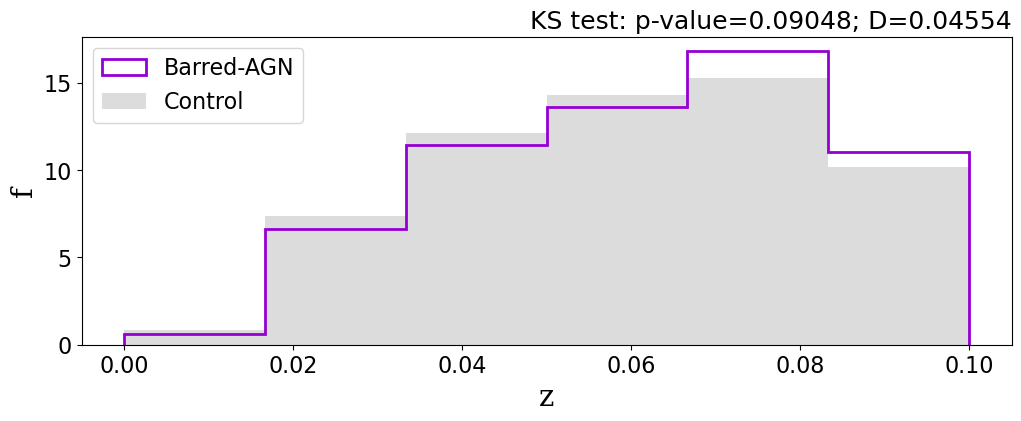}
   \includegraphics[scale=0.3]{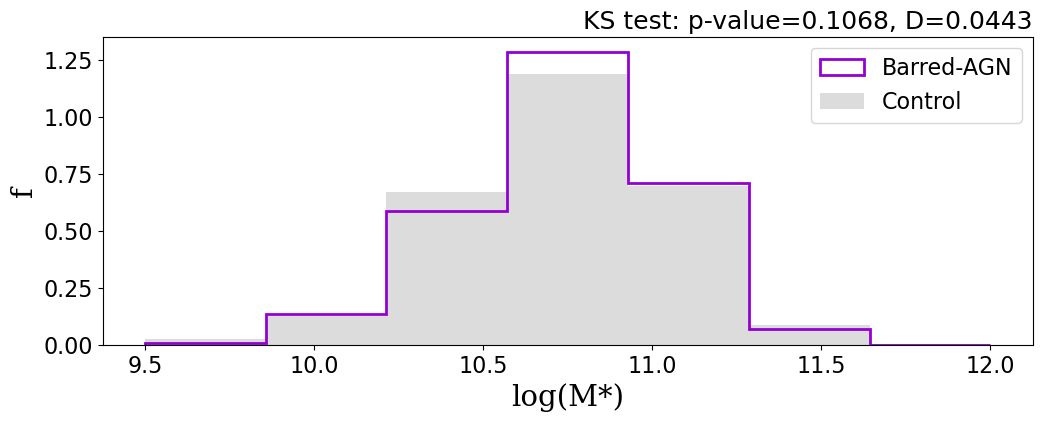}
   \includegraphics[scale=0.3]{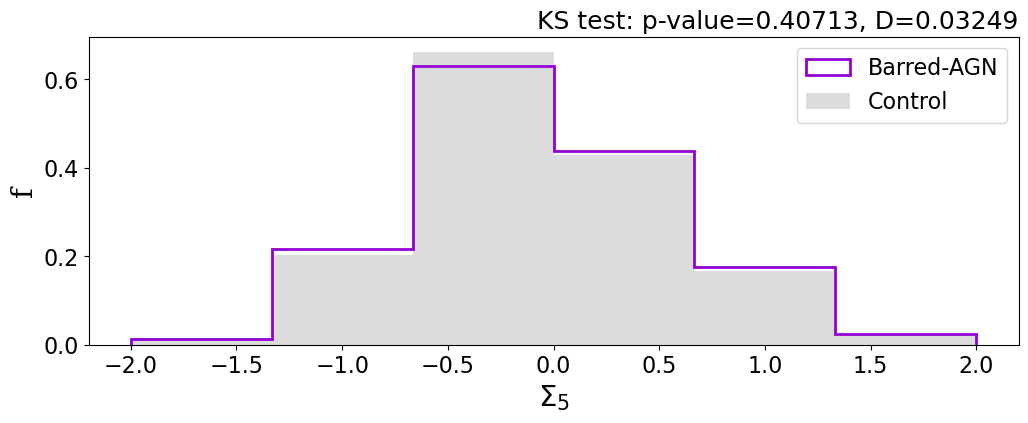}
   \includegraphics[scale=0.3]{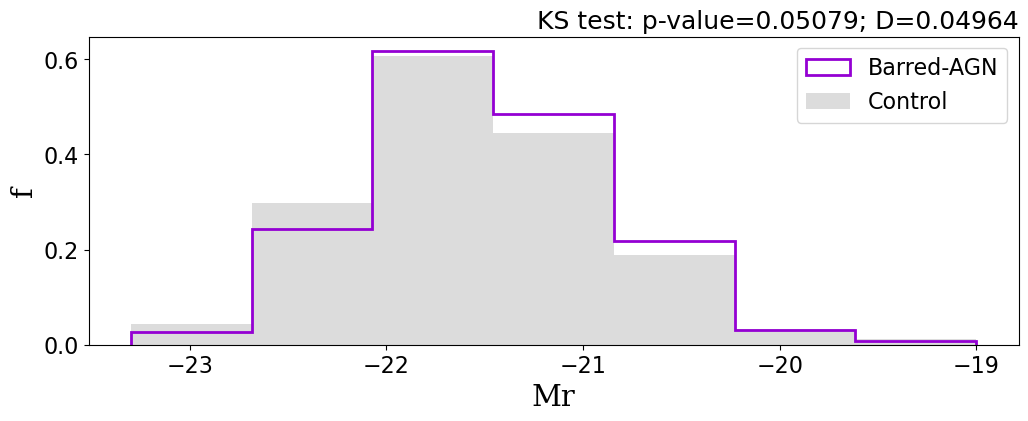}
   \includegraphics[scale=0.3]{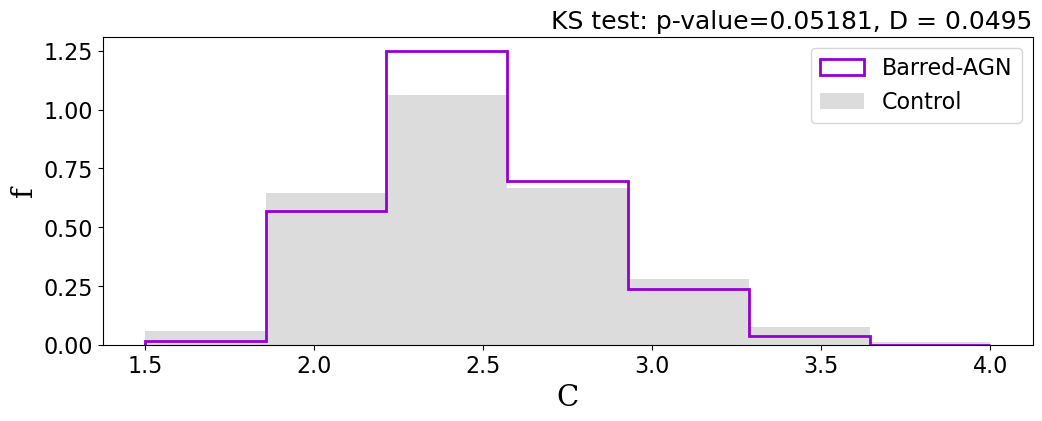}
   
      \caption{Normalized distributions of redshift, z, stellar mass, log($M^{\*}$),  the local density parameter $\Sigma_5$, absolute magnitude in the r band, $M_r$ and the concentration index $C$ for barred and unbarred galaxies. The D and p-values of the KS test are presented.}
         \label{control}
   \end{figure}

The statistical similarity between these parameters guarantees the validity of the comparison and the results obtained for the different properties of the host galaxy. This control sample of 1651 unbarred AGN galaxies, has a larger number of galaxies than the barred-AGN sample, allowing us to have confident statistical testing sets. 

This, together with similar distributions of the concentration index, which for high values is associated with bulge-like morphologies and for low values with spiral structure, allowed us to minimize the bias due to the presence of prominent bulges. In this way, we ensured that the results more accurately reflect the true impact of the bar on the host galaxy.

Furthermore, we performed a Kolmogorov Smirnov test \citep[KS;][]{Kstest} between the distribution of control and the main sample. From this test, we obtained a p value p$>$0.05 for the null hypothesis that the samples were drawn from the same distribution (see Fig. \ref{control}). The p-values of the KS test between the barred sample and CS are presented in the legends of Fig. \ref{control}.
In addition, the Anderson–Darling (AD, \cite{ADtest}) test returns p-values above 0.05 for the key parameters, providing further support for the consistency between the two samples and reinforcing the KS test results.
Table \ref{table1} presents the number of galaxies used for this work.
\\
  
 \begin{table}[h!]
      \caption[]{Catalogs obtained for this work}
         \label{table1}
     $$ 
         \begin{array}{p{0.5\linewidth}l}
            \hline
            \noalign{\smallskip}
            Sample &  Number \\
            \noalign{\smallskip}
            \hline
            \noalign{\smallskip}
            Barred AGN spiral galaxies & 1330 \\
            \\
            Non-barred AGN spiral galaxies (CS) & 1651 \\
            \noalign{\smallskip}
            \hline
         \end{array}
     $$
   \end{table}

\section{The influence of bars}
This work aims to analyze how the presence of bars affects the properties of AGN host galaxies and their role in SMBH activity, exploring their potential contribution to the mechanism that drives galaxy evolution. To investigate the influence of bars on the accretion-driven nuclear activity over the central black hole, we present a comparative study mainly focused in [OIII] $\lambda5007$ line, Lum[OIII], considered as a tracer of nuclear activity \citep{Kauffmann2003,Heckman2004}. It is one of the strongest narrow emission lines in this type of galaxies and its contamination by the star formation contribution is minimal, thus providing a reliable measure of nucleus power. To complement this study, we calculated the accretion rate of AGN, $\cal{R}$ \citep{Heckman2004}, an indirect estimator of the amount of material being accreted by the black hole given by the eq. \ref{eqR}. 

\begin{equation}
    \centering
    R= \frac{log(Lum[OIII])}{M_{BH}}
    \label{eqR}
\end{equation}

For this purpose, the mass of the black hole 
($M_{BH}$\footnote{$logM_{BH} = \alpha + \beta log(\frac{\sigma_*}{200})$}) was calculated for both barred and unbarred galaxies through the $M_{BH} - \sigma_*$ relation \citep{Ferrarese2000,Tremaine2002} where $\sigma_*$ is the bulge velocity dispersion. The velocity dispersion values were obtained from the MPA/JHU catalogues.
 
\cite{Graham2008} explain that the velocity dispersion may be affected due to the presence of a bar as it alters the stellar motion in the surroundings of the bulge so that the $M_{BH} - \sigma_*$ relation will be different for barred and non-barred galaxies.

\begin{figure}[h!]
   \centering
   \includegraphics[scale=0.43]{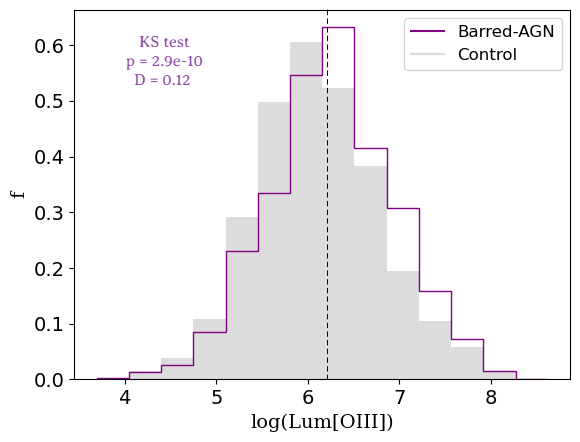}
   \includegraphics[scale=0.43]{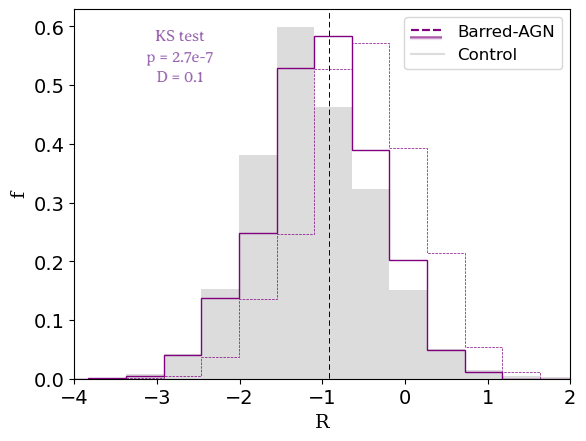}
      \caption{Normalized distributions of log(Lum[OIII]) (top panel) and the accretion rate $\cal{R}$ (bottom panel). The solid line represent the $\cal{R}$ distribution using the same $\alpha$ and $\beta$ parameter for both barred and unbarred galaxies. The dashed line represent the $\cal{R}$ distribution taking into account different values of alpha and beta for barred galaxies. The vertical line in both graphs represents the median of the barred sample.}
         \label{OIII-R}
   \end{figure}

Fig. \ref{OIII-R} shows the Lum[OIII] (top panel) and $\cal{R}$ (bottom panel) distributions for barred (solid line) and non-barred AGN (full surfaces). In the case of accretion rate, the dashed line represent the distribution of $\cal{R}$ using $\alpha=7.67\pm0.115$ and $\beta=4.08\pm0.751$. These parameters, as adopted in \cite{Alonso2018} illustrate that employing the specific values for barred galaxies result in an improved estimation of the accretion rate compared to those used for unbarred galaxies. However, to ensure a consistent comparison, the same $\alpha=8.13\pm0.06$ and $\beta=4.02\pm0.32$ \citep{Tremaine2002} values were applied to both barred (solid line) and unbarred (full surfaces) galaxies throughout the rest of the analysis.

In both plots it can be seen that AGN hosts exhibit differences between barred and non-barred galaxies, the barred galaxies show higher values of nuclear activity with respect to the control sample, however, this difference is less noticeable for the accretion rate. Compared to the barred sample, unbarred galaxies tend to show lower AGN luminosities  explained by the lack of there is less gas around the center, which will affect the influence of black hole feed. To quantify the differences between the barred AGN samples and their respective control samples, we performed KS-test on the distribution of log(Lum[OIII]) and the accretion rate R. The D and p-values are presented in Fig. \ref{OIII-R}

From these distributions, we separate between strong and weak AGNs subsamples. We consider the luminosity log(Lum[OIII]) = 6.2 as a threshold to differentiate powerful AGNs from weak ones, this limit represents a significant increase in nuclear activity and coincides with the median of the data. A similar approach was used by \cite{Alonso2018} to analyze in detail the excess of [OIII] luminosity and its relation with the presence of a bar. \cite{Kauffmann2003} considered powerful AGNs those with a log(Lum[OIII])$>$7, same limit used by \cite{Coldwell2009}.  In the case of accretion rate, we also define $\cal{R}$ = -0.9  as the value at which the excess becomes significant for the barred galaxies relative to the control sample, in the same way as for Lum[OIII], this value represents the median of the data. Table \ref{powerfulAGNs} quantifies the percentages of powerful AGNs selected according to thresholds in  Lum[OIII] and $\cal{R}$ respectively.

For the case of luminosity, the difference of barred galaxies with the control sample is shown for both the value selected by this paper and the one suggested by \cite{Kauffmann2003}, classified as extremely powerful. We can observe that barred galaxies tend to host more powerful AGNs compared to those without a bar.

\begin{table}[H]
    \centering
\begin{tabular}{ccc}
\hline
\noalign{\smallskip}
 & \% Barred-AGN & \% CS \\ 
\noalign{\smallskip}
\hline
\noalign{\smallskip}
Lum[OIII]$>10^{6.2}L_\odot$ & 53.4$\% \pm$ 1.3  & 42.3$\% \pm$ 1.2 \\ 
\noalign{\smallskip}
Lum[OIII]$>10^{7}L_\odot$ & 15$\% \pm$ 0.9 & 10$\% \pm$ 0.7\\ 
\noalign{\smallskip}
$\cal{R}>$-0.9 & 45$\% \pm$ 1.3 & 35$\% \pm$ 1.2 \\

\hline
\end{tabular}
 \caption{Percentages of AGN galaxies with a higher nuclear activity and extremely higher nuclear activity for the barred and control sample.}
 \label{powerfulAGNs}
\end{table}

\begin{figure*}
   \centering
   \includegraphics[scale=0.35]{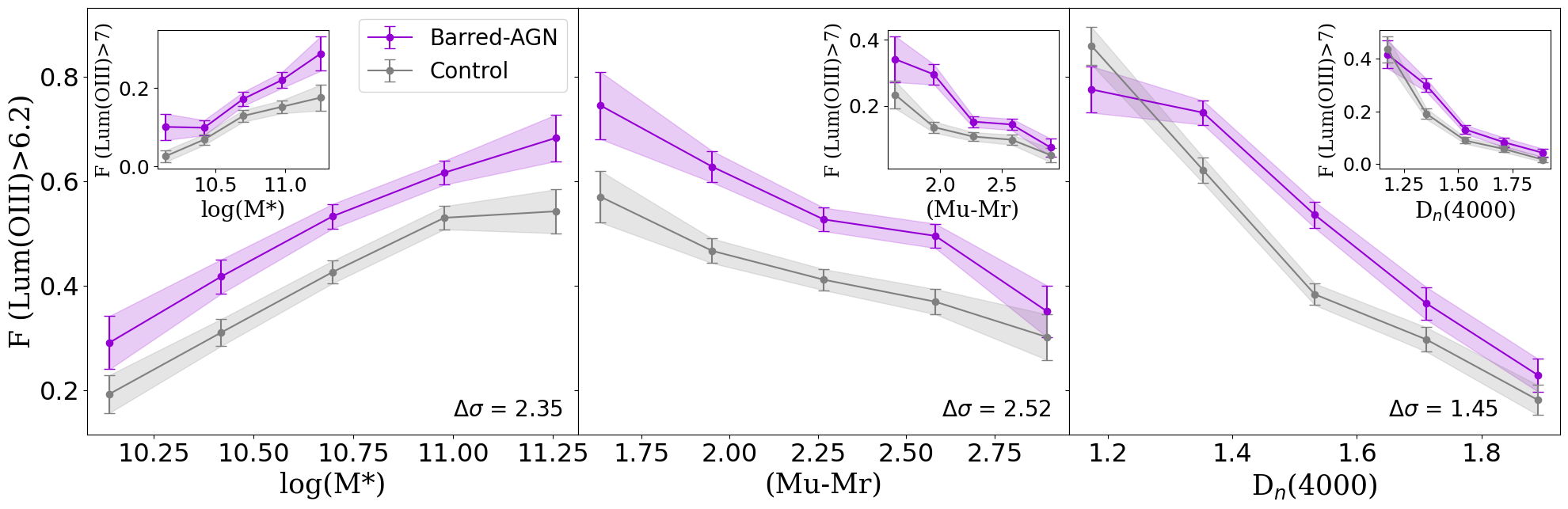}
   \caption{Fraction of Lum[OIII] $>$ $10^{6.2}L_{\odot}$ as a function of stellar mass, color (Mu-Mr), and the $D_n(4000)$ parameter (left, center, and right graphs, respectively). The values of the difference $\sigma$ between the samples in each of the panels are presented. The inner graphs show the fraction of Lum[OIII] $>$ $10^{7}L_{\odot}$ as a function of the same properties, this value is used by \cite{Kauffmann2003} as a measure of extremely excess Lum[OIII].}
    \label{LOIII}
    \end{figure*}

From this, we performed an analysis of the properties of host galaxies. In Fig \ref{LOIII} we present the fraction of galaxies with a powerful AGN with Lum[OIII]$>10^{6.2}$ as a function of stellar masses, (Mu-Mr) color and stellar age indicator $D_n(4000)$ for barred AGNs (violet lines) and unbarred AGNs (grey lines). The error bars were calculated using bootstrap error resampling \citep{Barrow1984} with 1000 iterations, applied to the calculation of the AGN fraction within each bin. 

The left panel shows the fraction of powerful AGNs as a function of stellar mass of the host galaxies. It is evident that those more massive host galaxies show a higher fraction of AGNs with an excess of luminosity, and within this, we also note a tendency of barred galaxies to present a higher fraction of powerful AGNs with respect to galaxies that do not have a barred structure regardless of stellar mass content.  These results are consistent with \cite{Alonso2018} and \cite{Oh2012}, where it is shown that the power of an AGN is moderately enhanced in the presence of a bar regardless of its stellar content. This increase is observed more strongly as we observe more massive galaxies.

In the middle panel of Fig. \ref{LOIII}, $Lum[OIII]>10^{6.2}$ fraction is analyzed as a function of $(Mu - Mr)$ color, there is a tendency for both samples to increase their fraction as the colors become bluer. Again, there is a uniform increase in the fraction of powerful AGNs for barred galaxies compared to unbarred galaxies, regardless of color. Similarly, \cite{Lee2012} found that AGN host galaxies in a certain color range ($2.0<(Mu-Mr)<2.5$) have higher activity in those with a bar.
Although these powerful AGNs exhibit bluer colors, generally the galaxies tend to have redder colors than non-barred galaxies \citep{Combes1981, Masters-N2010}. In our sample, approximately 57\% of barred galaxies have redder colors ((Mu-Mr)>2.3), compared with the 45\% of unbarred galaxies. The threshold Mu-Mr corresponds to the median of our color distribution and is consistent with values adopted in previous studies and the statements of both \cite{Alonso2013} and \cite{Oh2012}. The latter found that a significant number of barred galaxies are redder than other types of spiral galaxies being these at the blue peak. In addition \cite{Geron2021} using GZ DECaLS found that the strong bar fraction is higher in the red sequence and in quiescent galaxies. However, the color differences in barred galaxies may be due to several situations, \cite{Neumann2019} explains that there are three categories of bars: star-forming, non-star-forming and a third category of star-forming that fades into the center of the galaxy \citep{Verley2007} and is not possible to study due to AGN contamination. \\

Analyzing the simultaneous dynamical processes occurring in this type of galaxy is complex. In this context, it is essential to consider phenomena such as feedback, which can significantly influence the galactic environment and its evolution. AGN feedback acts as a dual regulatory mechanism that, depending on local and temporal conditions, can either compress the surrounding gas, promoting episodes of star formation at early stages \citep{Zubovas2014, Cresci2018}, or heat and expel the gas, suppressing star formation on longer timescales  \citep{Silk1998, Dimatteo2005, Fabian2012}.

In conjunction with bars, the gas flow induced by these structures not only intensifies nuclear activity, as suggested by the plots in Fig. \ref{LOIII}, but also amplifies the impact of AGN feedback. Observations and simulations indicate that bars facilitate the inward transport of cold gas, which may enhance star formation in central regions and contribute to AGN fueling \citep{Alonso2018, Lin2017, Vera2016, Robichaud2017}. The interplay between positive AGN feedback and the gas compression generated by the bars, which in turn sustains AGN activity, may extend star formation periods in the nuclear region. This could explain the bluer colors observed in galaxies hosting strong AGN, whether barred or unbarred \citep{Ellison2021}.

However, on longer timescales, negative feedback, combined with central gas depletion, may contribute to the transition of these galaxies to redder colors, aligning with broader galactic evolutionary processes \citep{Smethurst2016}. Some simulation-based studies, such as \cite{Hopkins2012}, suggest that AGN feedback may have a limited impact on the cold gas in the galactic nucleus, acting only on small amounts of gas. Nevertheless, more recent observational studies indicate that outflows driven by AGN can significantly affect star formation, even in gas-rich environments \citep{Fiore2017, Fluetsch2019}. These discrepancies highlight the complexity of the dynamical processes involved and underline the importance of bars as regulators in the redistribution of gas towards the nucleus.\\ 

In the right panel, $Lum[OIII]>10^{6.2}$ as a function of  spectral index $D_n(4000)$ is shown. This parameter is a spectroscopic measure defined as the average flux ratio between two narrow bands around the Balmer discontinuity at 4000 Angstroms, used as an indicator of the age of stellar populations. We observe a similar trend to the color, the fraction of galaxies with powerful AGNs tends to host younger stellar populations. In addition, barred galaxies show a slight enhancement in nuclear activity compared to the control sample. We associate this with the fact that galaxies with high nuclear activity are triggering episodes of star formation wich is in line with the findings of \cite{Ellison2011a}, who explains that central star formation rates are higher for barred galaxies than for unbarred ones, in a stellar mass range similar to the one used in this work. 

As a summary, we suggest that galaxies with a high fraction of powerful AGNs tend to be barred galaxies, which in turn tend to be massive, blue and with young stellar populations. 
For the purpose of corroborate the trends observed for each of the properties, the analysis was replicated this time for $Lum[OIII]>10^{7}$ (\ref{LOIII}, inner boxs), value used by \cite{Kauffmann2003} which represents extremely powerful AGNs. As can be seen, the trends remain similar to the limit change for all properties (stellar mass, $Mu - Mr$ color and $D_n$(4000) parameter), which implies stability in our results. In fact we also note that, for example, in the case of stellar mass, the tendency for galaxies with a high fraction of powerful AGNs becomes more pronounced as the galaxy becomes more massive. 

\begin{figure*}
   \centering
   \includegraphics[scale=0.35]{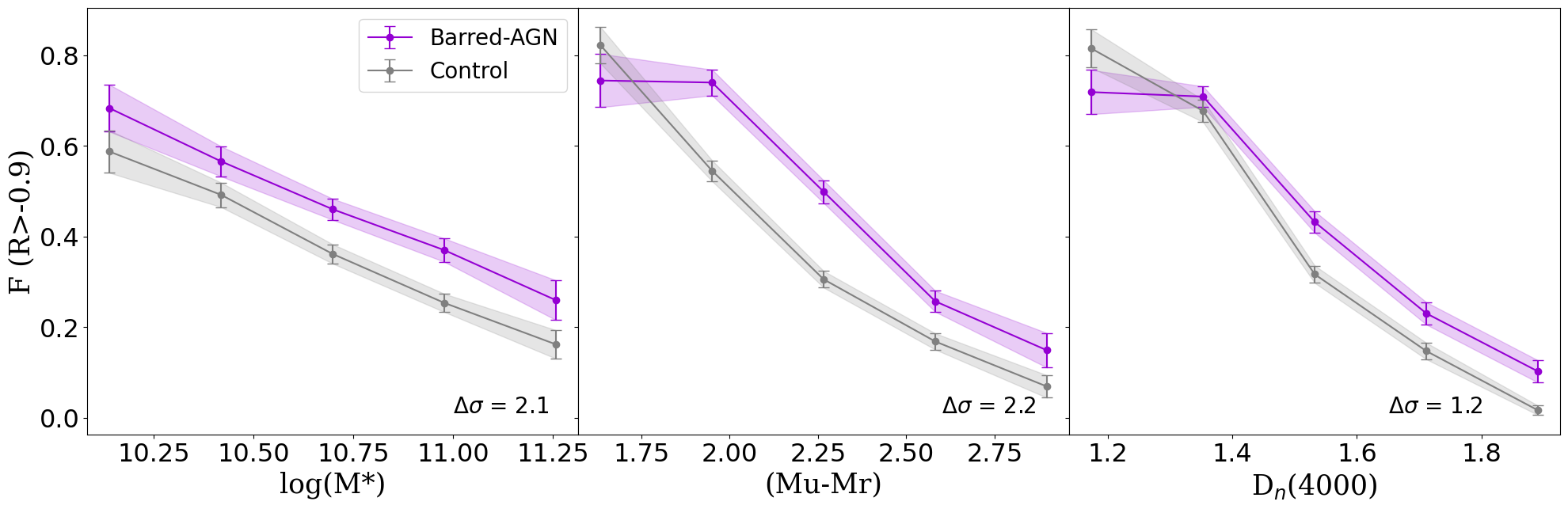}
   \caption{Fraction of $\cal{R}$ $>$ $-0.9$ as a function of stellar mass, color (Mu-Mr), and the $D_n(4000)$ parameter (left, center, and right graphs, respectively). The values of the difference $\sigma$ between the samples in each of the panels are presented. }
    \label{R}%
    \end{figure*}

In Fig. \ref{R} we present the fraction of galaxies with accretion rate $\cal{R}>$-0.9 as a function of the host properties (stellar mass, $Mu - Mr$ and $D_n(4000)$ parameter) with the aim of verifying in more detail the influence of the bar on SMBH activity. We note similar trends to those observed at [OIII] luminosity in both color and $D_n(4000)$, i.e., galaxies with higher nuclear activity (measured by the accretion rate parameter $\cal{R}$) tend to be bluer galaxies with younger stellar population for both barred and unbarred galaxies. However although an excess in the accretion rate is shown for the barred galaxies compared to the control sample, the range of difference is slightly less evident than at [OIII] luminosity. This is supported by the significance levels of the differences, quantified as $\Delta\sigma$ values labeled in each panel of plots for [OIII] and $\cal{R}$.

In the case of stellar mass, we observe an increase in the excess accretion rate for less massive barred and unbarred galaxies, with AGNs in barred host galaxies showing a moderate excess compared to their unbarred counterparts. Considering this result, it is important to assess the potential impact of the bar on the stellar mass of a galaxy. Generally, bars are not expected to significantly affect the total stellar mass of a galaxy but rather influence the redistribution of gas, thereby modulating central dynamics \citep{Athanassoula2003, Kormendy2004, Erwin2018}. However, simulations suggest that bars are more frequently found in less massive galaxies due to disk instabilities and dark matter halo distribution \citep{Athanassoula2013, Saha2018}.

To explain the observed trend, we propose that the relationship between bars and host galaxy mass is not necessarily causal. Instead, the structural and dynamical properties of galaxies shape the influence of bars, whose relative impact may be more pronounced at different mass scales \citep{Aguerri2023}. In less massive galaxies, bars could have a proportionally larger dynamical effect, as they can redistribute a significant fraction of gas and stars within a more compact volume. This increased efficiency in gas transport could enhance AGN fueling, potentially explaining the moderate accretion rate excess observed in barred AGNs.

Nevertheless, this result may be more closely linked to the available gas reservoir and black hole mass rather than solely to the presence of a bar. Given that lower-mass galaxies tend to host smaller black holes \citep{Greene2007, Reines2015}, the observed accretion rate excess may reflect the interplay between the bar-driven gas inflow and the relative scale of the black hole's gravitational influence. Further analysis is needed to disentangle the contributions of bars, gas availability, and black hole scaling relations in driving these trends.

\begin{figure}[h!]
   \centering
   \includegraphics[scale=0.37]{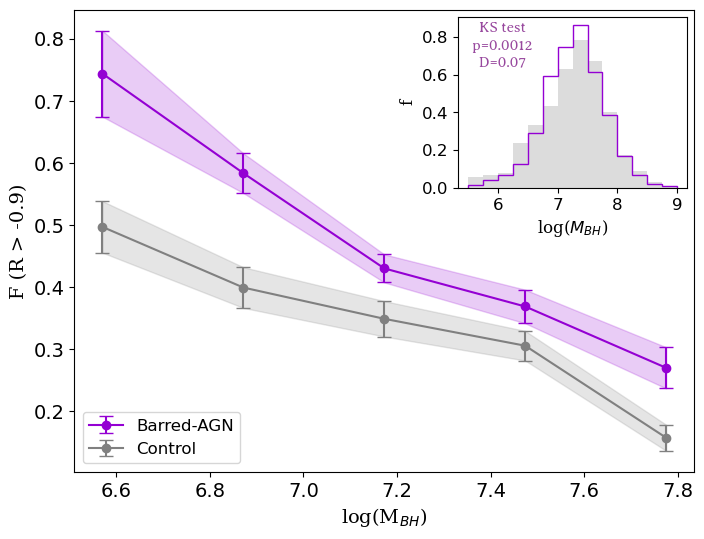}
      \caption{Fraction of $\cal{R}$ $>$ $-0.9$ as a function of black hole mass. Inner box show the distribution of black hole mass and values of KS-test are presented.}
         \label{R-Mbh}
   \end{figure}

The Fig.\ref{R-Mbh} shows the fraction of powerful AGNs ($\cal{R}>$-0.9) as a function of black hole mass. We notice that the fraction of galaxies with high nuclear activity increases for lower mass black holes, showing that smaller black holes apparently accrete matter with greater efficiency. In addition, this trend is more pronounced for barred galaxies. The inner plot shows the black hole mass distribution for both barred galaxies and the control sample. There is a slight tendency for barred galaxies to host smaller black holes. 

Scaling relations ($M_{BH}-M_{bulge}$, $M_{BH}-\sigma$), provide robust evidence for the coevolution of host galaxy and supermassive black holes, with AGN feedback playing a key role in shaping this correlation \citep{mcconnell2013}. In conjunction with our results, this motivates an analysis of the structural properties of the host galaxies. Barred galaxies are frequently associated with pseudobulges, which form through secular processes such as gas redistribution and sustained star formation \citep{kormendy2013}. Unlike classical bulges, which are typically the result of rapid mergers and violent relaxation, pseudobulges grow gradually through internal secular evolution. As a result, they generally host smaller black holes compared to classical bulges, which could explain the higher fraction lower-mass black holes in barred galaxies.   
Furthermore, according to our results smaller black holes are expected to accrete more efficiently relative to their mass, especially during early stages of their growth as reported on \cite{Alexander2012, Greene2007}.
This enhanced accretion efficiency, combined with the ability of bars to drive gas inflows toward the galactic center, suggests that bars influence the internal dynamics that mediate AGN fueling, potentially leading to more active nuclear regions.

\subsection{Types of bars} 
\label{Types_bars}

Bars are typically classified into strong or weak (strongly barred, SB, and weakly barred, SAB) depending on its prominence or light distribution and ability to redistribute mass and angular momentum in the galaxy \citep{Vaucouleurs1963}. Visual classification remains a widely used method, as demonstrated by \cite{Nair2010}, who constructed a catalog of galaxies in which bars were classified as strong or weak depending on whether they dominated the light distribution. Other studies, such as \cite{Gadotti2011}, have taken a quantitative approach by estimating bar strength based on the ellipticity of isophotes.
 Recently, \cite{Geron2021, Geron2023} used Galaxy Zoo DECaLS to study strong and weak bars in disk galaxies. They found that around 28\% of all disk galaxies have a weak bar and a 16\% had a strong bar, also found that are not fundamentally different physical phenomena between this classifications and suggest that exist a continuum of bar type. On the other hand, multiple studies have suggested that there are differences in the surface brightness profile of weak and strong bars \citep{Elmegreen1996, Kim2015}.

    \begin{figure}[h!]
   \centering
   \includegraphics[scale=0.44]{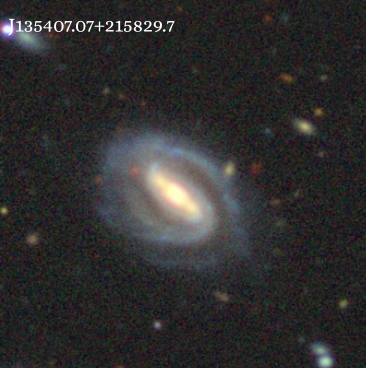}
   \includegraphics[scale=0.44]{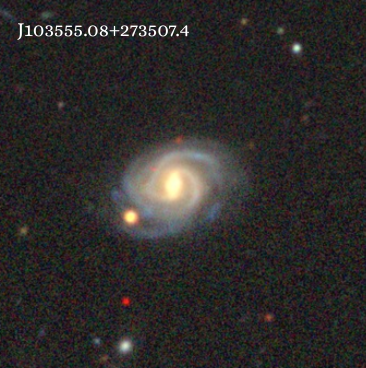}
      \includegraphics[scale=0.44]{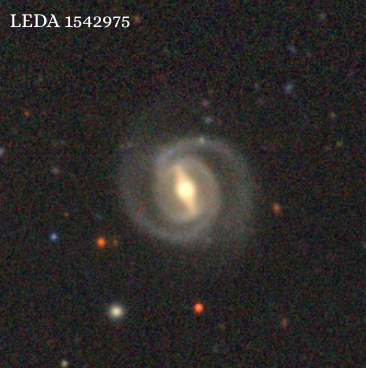}
   \includegraphics[scale=0.44]{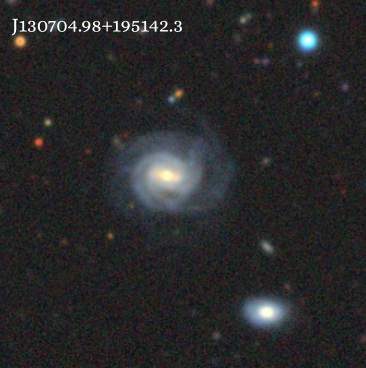}
      \caption{Example of strong barred galaxies (left panel) and weak barred galaxies (right panel) from GZ DECaLS.}
         \label{example2}
   \end{figure}
   
For this study, we use the morphological classification provided by GZ DECaLS to differentiate between strong and weak bars in our sample of barred AGNs (see Section \ref{agn_sample}). Fig \ref{example2} presents typical examples of galaxies hosting strong and weak bars, while Table \ref{table-strong-weak} summarizes the percentage of AGNs classified as strongly or weakly barred within our sample.

 \begin{table}[h!]
      \caption[]{Numbers and percentages of strong and weak barred AGNs}
         \label{table-strong-weak}
     $$ 
         \begin{array}{p{0.37\linewidth}l l}
            \hline
            \noalign{\smallskip}
            Sample & Number & Percentage \\
            \noalign{\smallskip}
            \hline
            \noalign{\smallskip}
            Strong-barred AGN & 865 &  65\% \\
            \noalign{\smallskip}
            Weak-barred AGN & 465 &  35\% \\
            \noalign{\smallskip}
            \hline
           \end{array}
     $$
   \end{table}

  \begin{figure*}[h!]
   \centering
   \includegraphics[scale=0.35]{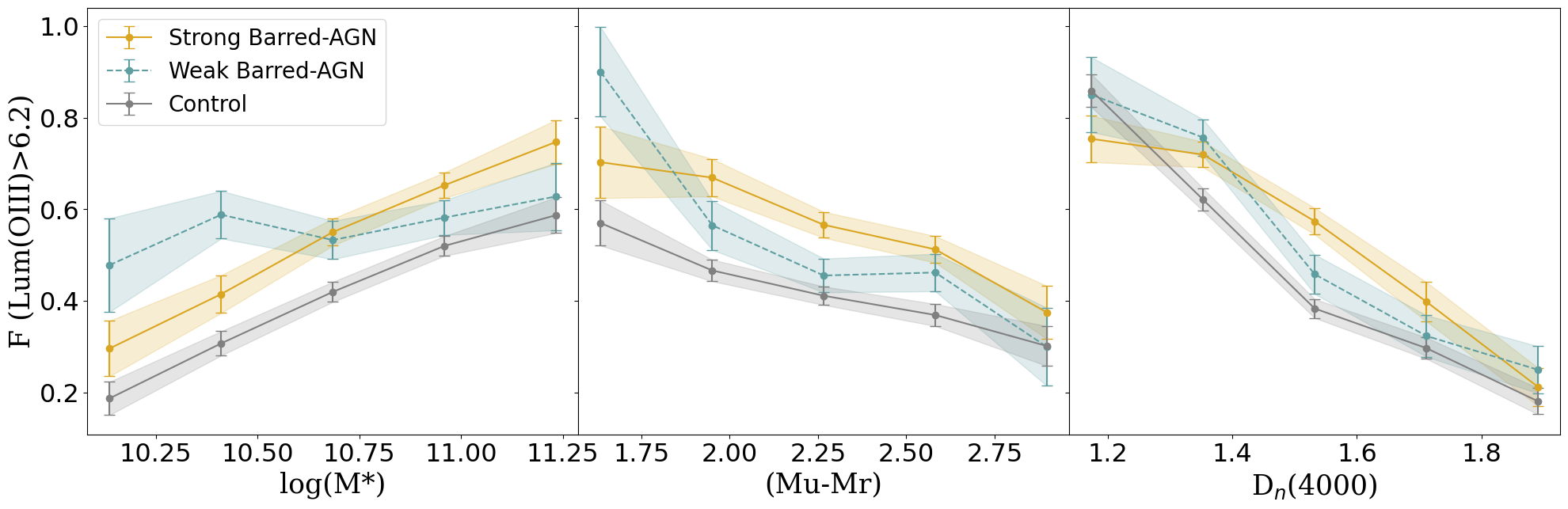}
   \caption{Fraction of Lum[OIII] $>$ $10^{6.2}L_{\odot}$ as a function of stellar mass, color (Mu-Mr), and the $D_n(4000)$ parameter (left, center, and right graphs, respectively). The graph separates by bar strength, either strong (yellow line) or weak (lightblue dashed line), compared to the control sample.}
    \label{OIII strong-weak}
    \end{figure*}

    \begin{figure*}[h!]
   \centering
   \includegraphics[scale=0.35]{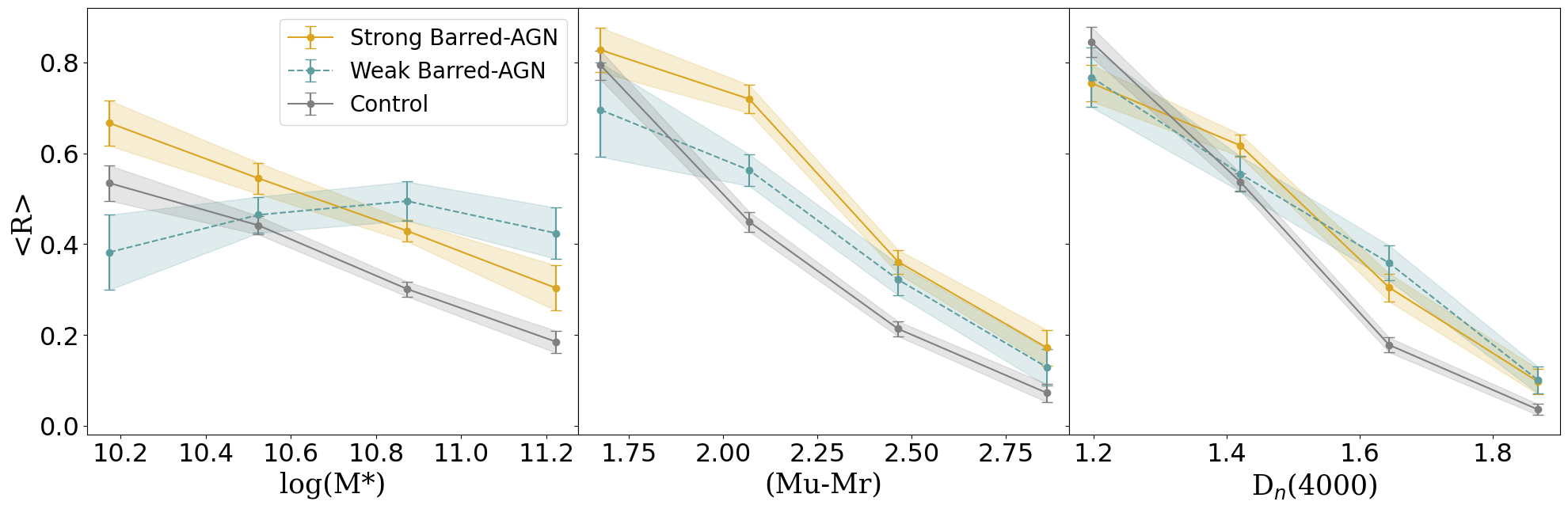}
   \caption{Fraction of $\cal{R}$ $>$ $-0.9$ as a function of stellar mass, color (Mu-Mr), and the $D_n(4000)$ parameter (left, center, and right graphs, respectively). The graph separates by bar strength, either strong (yellow line) or weak (lightblue dashed line), compared to the control sample.}
    \label{R strong-weak}
    \end{figure*}

Fig. \ref{OIII strong-weak} and \ref{R strong-weak} show the fraction of galaxies hosting powerful AGNs as a function of host galaxies properties, this time distinguishing between strong bars (yellow line) and weak bars (light blue line). 
We observe that galaxies with strong bars follow the main trend seen in Fig. \ref{LOIII} and Fig.\ref{R}, whereas galaxies with weak bars exhibit greater dispersion. This observed trend suggest that strong bars could have a more significant role in AGN fueling, potentially leading to a more consistent pattern. However, despite the observed trend in strongly barred galaxies, we do not find a substantial increase in the fraction of powerful AGNs compared to weakly barred ones. This indicates that while bar strength may play a role in AGN fueling, additional factors likely contribute to the observed distribution.

Nevertheless, given the limitations imposed by the sample size, this result should be interpreted with caution, as the smaller number of weakly barred galaxies introduces an additional source of uncertainty. A more robust statistical evaluation, incorporating a larger sample of weakly barred galaxies, would be necessary to confirm this trend.

Several studies suggest that strong bars have a significant impact on their host galaxies. For instance, \cite{Vera2016} found that strongly barred galaxies tend to be redder than weakly barred ones and exhibit much lower star formation efficiency. This implies that prominent bars play a crucial role in regulating gas processing within the galaxy.

From a different perspective, \cite{Garland2024} reported that galaxies with strong bars are more likely to host an AGN than those with weak bars, which is consistent with the results of this work. This finding suggests that strong bars may be more effective in triggering or enhancing black hole activity, potentially due to their ability to drive gas inflows more efficiently toward the central regions.

If strong bars are indeed more effective at driving gas inflows, they could play a key role in shaping the co-evolution of supermassive black holes and their host galaxies. However, as \cite{Geron2023} suggested, bar strength exists along a continuum rather than discrete categories, implying that the impact of bars on AGN activity may vary gradually rather than abruptly. 

In summary, based on our results and previous evidence of differences between strong and weak bars, we propose that strong bars may have a more significant impact on nuclear activity compared to less prominent bars (see Table \ref{table2}). 
However, future studies with larger statistical samples and multi-wavelength observations will be essential to further explore the link between bar strength, AGN fueling efficiency, and the long-term evolution of disk galaxies.

\begin{table}[H]
    \centering
    
\begin{tabular}{ccc}
\hline
\noalign{\smallskip}
 & $\%$ Strong-Barred AGN & $\%$ Weak-Barred AGN \\ 
\noalign{\smallskip}
\hline
\noalign{\smallskip}
$Lum[OIII]>10^{6.2}L_\odot$ & $55.9\% \pm 1.6$  & $48.6\% \pm 2.3$  \\ 
\noalign{\smallskip}
$\cal{R}$>-0.9 & $47.0\% \pm 1.6$  & $40.2\%\pm 2.2$  \\
   
\hline
\end{tabular}
 \caption{Percentages of strong and weak barred AGN galaxies with powerful nuclear activity}
 \label{table2}
\end{table}

\subsection{The impact of environment}  

Galaxy morphology is known to correlate with environmental density, with spiral galaxies preferentially inhabiting lower-density regions, while elliptical galaxies are more commonly found in dense environments \citep{Dressler1980}. However, the influence of the local environment on the presence of bars remains a topic of debate.

Several observational studies suggest a correlation between barred galaxies and denser environments. \cite{Eskridge2000,Skibba2012} and \cite{Alonso2014} reported a higher fraction of barred galaxies in dense regions, suggesting that external factors may influence bar formation or longevity. Numerical simulations further support this idea, indicating that bars can be induced by galaxy interactions \citep{Mihos1997,Berentzen2004}. Additionally, \cite{Elmegreen1990} found a higher fraction of barred galaxies in galaxy pairs, reinforcing the notion that interactions may play a role in bar formation. Conversely, other studies found no significant dependence of bar frequency on environmental density. \cite{Mendez2010, Lee2012} and \cite{Martinez2011} did not observe substantial differences in the fraction of barred galaxies between field and cluster environments, suggesting that bars may primarily form through internal secular processes rather than external influences.

This lack of consensus highlights the complexity of bar formation and evolution, which may depend not only on local density but also on other structural and dynamical properties of galaxies.

Motivated by this, in this work, we analyze the fraction of galaxies with strong bars, weak bars and no bars as a function of their galactic environment, differentiating between galaxies dense, medium and isolated environments. \cite{Alonso2006} stablish three regions to classify environments according to this parameter: low density (log$(\Sigma_5)<$-0.57), medium density (-0.57 $<$ log$(\Sigma_5)<$0.05) and high density (log$(\Sigma_5)>$0.05). Since our sample is designed to maintain a consistent environmental distribution between barred and unbarred galaxies, we can assess the effect of the bar on AGN activity without biases introduced by environmental factors. 

\begin{figure}[h!]
    \centering
    \includegraphics[scale=0.6]{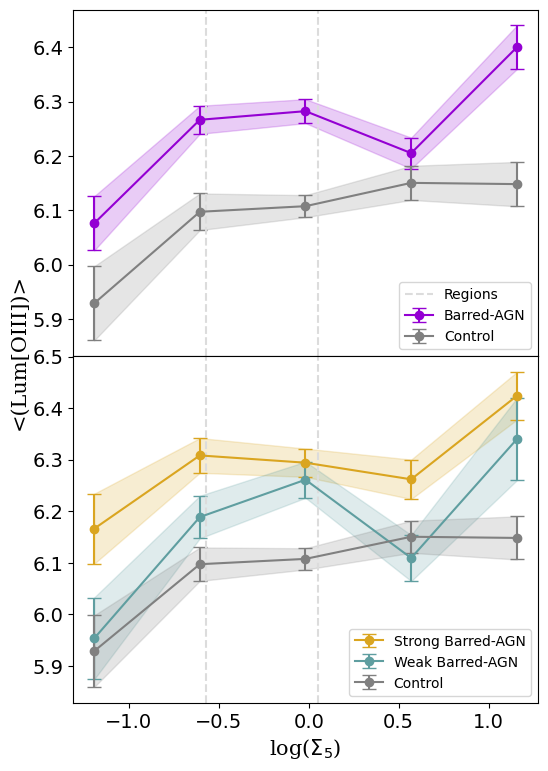}
    \caption{<Lum[OIII]> of AGN as a function of log($\Sigma_5$) for full barred sample (top panel, violet line) and the control sample (grey line.) and the sample separated (bottom panel) into strong bars (yellow line) and weak bars (lightblue line) in addition to the control sample. } 
    \label{lo3vssigma}
\end{figure}

Fig. \ref{lo3vssigma} presents the mean [OIII] luminosity as a function of the local density parameter. In both panels, a slight trend is observed, indicating that galaxies tend to exhibit higher Lum[OIII] values in denser environment. Focusing on the bottom plot, it is evident that galaxies with strong bars follow the overall trend while maintaining higher Lum[OIII] values, whereas those with weak bars display a more scattered distribution.

\begin{table*}
\center
\caption{Percentages of samples with higher values of Lum[OIII] and $\cal{R}$ for barred and non-barred galaxies in different environments.}
\begin{tabular}{c c c c c}
\hline
& \multicolumn{2}{c}{Barred Sample} & {Control Sample}   \\
\hline
  $Lum[OIII]>6.2$ & Strong bars $\%$ &  Weak bars $\%$ & Without bar \\  
\hline
\hline
 Environment & &  & &\\
\hline
 Low density      & 54.4 $\pm$ 3.7  &  46.5 $\pm$ 4.9  &  39.5 $\pm$ 2.7 \\
\hline
  Medium density  & 55.5 $\pm$ 3.6 &  52.3 $\pm$ 3.5  &  42.0 $\pm$ 1.8  \\
\hline
  High density    & 57.1 $\pm$ 2.6  &  45.5 $\pm$ 3.8  &  43.9 $\pm$ 1.9 \\
\hline
\end{tabular}
\vspace{0.5cm}

\begin{tabular}{c c c c c}

\hline
  $\cal{R}$>-0.9 & Strong bars $\%$ &  Weak bars $\%$ & Without bar \\  
\hline
\hline
 Environment & &  & &\\
\hline
 Low density      & 48.8 $\pm$ 3.7  &  29.7 $\pm$ 4.5  &  38.3 $\pm$ 2.7 \\
\hline
  Medium density  & 46.9 $\pm$ 2.7  &  47.6 $\pm$ 3.5  &  36.5 $\pm$ 1.8 \\
\hline
  High density    & 46.2 $\pm$ 2.6  &  37.8 $\pm$ 3.7 &  32.7 $\pm$ 1.8 \\
\hline
\end{tabular}

{\small  }
\label{tablaresumen}
\end{table*}

To quantify this trend, Table \ref{tablaresumen} shows the percentages of barred galaxies and the control sample with high nuclear activity separated by bar strength in the case of the barred sample and also in different environments. Our results indicate that in intermediate-density environments, the fraction of powerful AGNs does not exhibit significant differences between galaxies with strong and weak bars. However, at the extremes of the environmental distribution, i.e. -low dense and very dense environments- these differences become more pronounced.

It is possible that the explanation for this result is mostly linked to environmental effects that regulate the availability of gas. In dense environments, interactions with the intracluster medium (ICM) and gravitational perturbations from nearby galaxies can enhance the inflow of bar-driven gas, increasing its strength. However, in low density environments, evolution is primarily secular, meaning it is governed by internal processes over long timescales. If a strong bar forms early and remains stable, it has more time to efficiently channel gas toward the central region, facilitating recurrent episodes of AGN activity.  

\section{Summary and conclusions}

We performed a comparative analysis to determine the impact or influence that a bar can have on the nuclear activity of a galaxy. For this purpose, we consider the sample of AGNs constructed by \cite{Coldwell2017} and a sample of barred galaxies classified by Galaxy Zoo DECaLS to obtain a final sample of spiral barred AGNs. In order to obtain a reliable measure of this comparison, we constructed a suitable control sample of unbarred AGNs galaxies, with similar characteristics in redshift, stellar mass, absolute magnitude, concentration index and local environment distributions. 
The use of Galaxy Zoo DECaLS in this study has been critical due to the depth and quality of the Dark Energy Camera Legacy Survey images, combined with morphological classifications provided by thousand of participants through citizen science in combination with deep learning. This approach has provided access to a large and diverse dataset, with broad coverage in terms of redshift and magnitude, facilitating the analysis of morphological properties at a level of detail that is difficult to achieve.

We can summarize the main results of this analysis in the following conclusions: 

\begin{itemize}

    \item We find that barred galaxies exhibit an overall enhancement of nuclear activity compared to their unbarred counterparts. Using a threshold of $Lum[OIII]>6.2$ to distinguish the powerful AGNs from the weaker ones, we analyzed the fraction of powerful AGNs as a function of stellar mass, color and $D_n(4000)$ index. Our results indicate that galaxies with higher nuclear activity tend to be more massive, bluer and host younger stellar population. Notably barred galaxies show an excess of Lum[OIII] compared to the control sample. 

    \item We also perform an accretion rate ($\cal{R}$) analysis, which reveals that the excess accretion rate is more pronounced in less massive galaxies, with barred AGNs exhibiting a moderate enhancement compared to the unbarred sample. We suggest that in low-mass galaxies, bars may have a proportionally larger dynamical effect, redistributing a significant fraction of gas and stars within a more compact region. This gas transport efficiency could enhance AGN fueling, providing a plausible explanation for the excess accretion rate observed in barred AGNs at lower stellar masses. Furthermore, we found that barred galaxies tend to host smaller black holes, which aligns with the observed trends.

    \item Based on the classifications provided by Galaxy Zoo, we explore the increase in nuclear activity by differentiating between strong and weak bars. We observe a slight tendency for strong bars to show higher values of nuclear activity than weak bars.

    \item We also analyzed the impact the environment in the AGN activity. Consistent with the idea that bars show no preference for specific environment, we noticed variations in AGN power as a function of bar strength in different environments. Specifically, in the most extreme environments-wether low or high density regions-galaxies with strong bars tend to show higher mean [OIII] luminosity than those with weak bars, but this tendency is not observed in intermediate-dense environment.

\end{itemize}

This study suggests that bars serve as important regulatory factors, potentially increasing nuclear activity and playing a significant role in the AGN phenomenon, contributing to the internal redistribution of gas, further facilitating AGN fueling. However, they may not be the primary determinant, as other factors such as gas availability, bar stability, and the evolutionary state of the SMBH could also influence AGN activity.

\begin{acknowledgements}
      This work was partially supported by DIDULS/ULS, through the projects PR2353857, PTE23538516 and PTE23538510 and also by Consejo Nacional de Investigaciones Cient\'{\i}ficas y T\'ecnicas and the Secretar\'{\i}a de Ciencia y T\'ecnica de la Universidad Nacional de San Juan. 
      Funding for the SDSS has been provided by the Alfred P. Sloan Foundation, the Participating Institutions, the National Science Foundation, the U.S. Department of Energy, the National Aeronautics and Space Administration, the Japanese Monbukagakusho, the Max Planck Society, and the Higher Education Funding Council for England. The SDSS Web Site is http://www.sdss.org/.
The SDSS is managed by the Astrophysical Research Consortium for the Participating Institutions. The Participating Institutions are the American Museum of Natural History, Astrophysical Institute Potsdam, University of
Basel, University of Cambridge, Case Western Reserve University, University of Chicago, Drexel University, Fermilab, the Institute for Advanced Study, the Japan Participation Group, Johns Hopkins University, the Joint Institute for Nuclear Astrophysics, the Kavli Institute for Particle Astrophysics and Cosmology, the Korean Scientist Group, the Chinese Academy of Sciences (LAMOST), Los Alamos National Laboratory, the Max-Planck-Institute for Astronomy (MPIA), the Max-Planck-Institute for Astrophysics (MPA), New Mexico State University, Ohio State University, University of Pittsburgh, University of Portsmouth, Princeton University, the United States Naval Observatory, and the University of Washington.
The Legacy Surveys consist of three individual and complementary projects: the Dark Energy Camera Legacy Survey (DECaLS; Proposal ID $\#$2014B-0404; PIs: David Schlegel and Arjun Dey), the Beijing-Arizona Sky Survey (BASS; NOAO Prop. ID $\#$2015A-0801; PIs: Zhou Xu and Xiaohui Fan), and the Mayall z-band Legacy Survey (MzLS; Prop. ID $\#$2016A-0453; PI: Arjun Dey). DECaLS, BASS and MzLS together include data obtained, respectively, at the Blanco telescope, Cerro Tololo Inter-American Observatory, NSF’s NOIRLab; the Bok telescope, Steward Observatory, University of Arizona; and the Mayall telescope, Kitt Peak National Observatory, NOIRLab. Pipeline processing and analyses of the data were supported by NOIRLab and the Lawrence Berkeley National Laboratory (LBNL). The Legacy Surveys project is honored to be permitted to conduct astronomical research on Iolkam Du’ag (Kitt Peak), a mountain with particular significance to the Tohono O’odham Nation.
NOIRLab is operated by the Association of Universities for Research in Astronomy (AURA) under a cooperative agreement with the National Science Foundation. LBNL is managed by the Regents of the University of California under contract to the U.S. Department of Energy.
This project used data obtained with the Dark Energy Camera (DECam), which was constructed by the Dark Energy Survey (DES) collaboration. Funding for the DES Projects has been provided by the U.S. Department of Energy, the U.S. National Science Foundation, the Ministry of Science and Education of Spain, the Science and Technology Facilities Council of the United Kingdom, the Higher Education Funding Council for England, the National Center for Supercomputing Applications at the University of Illinois at Urbana-Champaign, the Kavli Institute of Cosmological Physics at the University of Chicago, Center for Cosmology and Astro-Particle Physics at the Ohio State University, the Mitchell Institute for Fundamental Physics and Astronomy at Texas $A\&M$ University, Financiadora de Estudos e Projetos, Fundacao Carlos Chagas Filho de Amparo, Financiadora de Estudos e Projetos, Fundacao Carlos Chagas Filho de Amparo a Pesquisa do Estado do Rio de Janeiro, Conselho Nacional de Desenvolvimento Cientifico e Tecnologico and the Ministerio da Ciencia, Tecnologia e Inovacao, the Deutsche Forschungsgemeinschaft and the Collaborating Institutions in the Dark Energy Survey. The Collaborating Institutions are Argonne National Laboratory, the University of California at Santa Cruz, the University of Cambridge, Centro de Investigaciones Energeticas, Medioambientales y Tecnologicas-Madrid, the University of Chicago, University College London, the DES-Brazil Consortium, the University of Edinburgh, the Eidgenossische Technische Hochschule (ETH) Zurich, Fermi National Accelerator Laboratory, the University of Illinois at Urbana-Champaign, the Institut de Ciencies de l’Espai (IEEC/CSIC), the Institut de Fisica d’Altes Energies, Lawrence Berkeley National Laboratory, the Ludwig Maximilians Universitat Munchen and the associated Excellence Cluster Universe, the University of Michigan, NSF’s NOIRLab, the University of Nottingham, the Ohio State University, the University of Pennsylvania, the University of Portsmouth, SLAC National Accelerator Laboratory, Stanford University, the University of Sussex, and Texas $A\&M$ University.
BASS is a key project of the Telescope Access Program (TAP), which has been funded by the National Astronomical Observatories of China, the Chinese Academy of Sciences (the Strategic Priority Research Program “The Emergence of Cosmological Structures” Grant $\#$ XDB09000000), and the Special Fund for Astronomy from the Ministry of Finance. The BASS is also supported by the External Cooperation Program of Chinese Academy of Sciences (Grant $\#$ 114A11KYSB20160057), and Chinese National Natural Science Foundation (Grant $\#$ 12120101003, $\#$ 11433005).
The Legacy Survey team makes use of data products from the Near-Earth Object Wide-field Infrared Survey Explorer (NEOWISE), which is a project of the Jet Propulsion Laboratory/California Institute of Technology. NEOWISE is funded by the National Aeronautics and Space Administration.
The Legacy Surveys imaging of the DESI footprint is supported by the Director, Office of Science, Office of High Energy Physics of the U.S. Department of Energy under Contract No. DE-AC02-05CH1123, by the National Energy Research Scientific Computing Center, a DOE Office of Science User Facility under the same contract; and by the U.S. National Science Foundation, Division of Astronomical Sciences under Contract No. AST-0950945 to NOAO.
\end{acknowledgements}

\bibliographystyle{aasjournal} 
\bibliography{bibliografia}

\end{document}